\documentclass[sigconf]{acmart}

\AtBeginDocument{%
\providecommand\BibTeX{{\normalfont B\kern-0.5em{\scshape i\kern-0.25em b}\kern-0.8em\TeX}}}

\hypersetup{draft}
\usepackage[T1]{fontenc}
\usepackage{enumitem}
\usepackage{afterpage}
\usepackage{listings}
\usepackage{xcolor}
\usepackage{graphicx}
\usepackage{caption}
\usepackage{placeins}
\usepackage{url}
\usepackage{epstopdf}
\usepackage{booktabs}
\usepackage{footnote}
\usepackage{hyphenat}
\makesavenoteenv{table}
\usepackage[acronyms,nonumberlist,nopostdot,nomain,nogroupskip]{glossaries}
\usepackage{subfig}
\usepackage{balance}

\newacronym{quic}{QUIC}{Quick UDP Internet Connections}
\newacronym{3gpp}{3GPP}{3rd Generation Partnership Project}
\newacronym{adc}{ADC}{Analog to Digital Converter}
\newacronym{5g}{5G}{5th generation}
\newacronym{aimd}{AIMD}{Additive Increase Multiplicative Decrease}
\newacronym{am}{AM}{Acknowledged Mode}
\newacronym{amc}{AMC}{Adaptive Modulation and Coding}
\newacronym{aqm}{AQM}{Active Queue Management}
\newacronym{awgn}{AGWN}{Additive White Gaussian Noise}
\newacronym{balia}{BALIA}{Balanced Link Adaptation}
\newacronym{bdp}{BDP}{Bandwidth-Delay Product}
\newacronym{bf}{BF}{Beamforming}
\newacronym{cc}{CC}{Congestion Control}
\newacronym{cdf}{CDF}{Cumulative Distribution Function}
\newacronym{cn}{CN}{Core Network}
\newacronym{cqi}{CQI}{Channel Quality Information}
\newacronym{cp}{CP}{Control Plane}
\newacronym{csirs}{CSI-RS}{Channel State Information - Reference Signal}
\newacronym{dc}{DC}{Dual Connectivity}
\newacronym{dce}{DCE}{Direct Code Execution}
\newacronym{dci}{DCI}{Downlink Control Information}
\newacronym{dl}{DL}{Downlink}
\newacronym{dmr}{DMR}{Deadline Miss Ratio}
\newacronym{dmrs}{DMRS}{DeModulation Reference Signal}
\newacronym{e2e}{E2E}{End-to-End}
\newacronym{ecn}{ECN}{Explicit Congestion Notification}
\newacronym{edf}{EDF}{Earliest Deadline First}
\newacronym{enb}{eNB}{evolved Node Base}
\newacronym{epc}{EPC}{Evolved Packet Core}
\newacronym{es}{ES}{Edge Server}
\newacronym{fdma}{FDMA}{Frequency Division Multiple Access}
\newacronym{fdd}{FDD}{Frequency Division Duplexing}
\newacronym[firstplural=Radio Access Technologies (RATs)]{rat}{RAT}{Radio Access Technology}
\newacronym{fs}{FS}{Fast Switching}
\newacronym{ftp}{FTP}{File Transfer Protocol}
\newacronym{gnb}{gNB}{Next Generation Node B}
\newacronym{harq}{HARQ}{Hybrid Automatic Repeat reQuest}
\newacronym{hetnet}{HetNet}{Heterogeneous Network}
\newacronym{hh}{HH}{Hard Handover}
\newacronym{hol}{HOL}{Head-of-Line}
\newacronym{ia}{IA}{Initial Access}
\newacronym{imt}{IMT}{International Mobile Telecommunication}
\newacronym{iot}{IoT}{Internet of Things}
\newacronym{los}{LOS}{Line of Sight}
\newacronym{lte}{LTE}{Long Term Evolution}
\newacronym{m2m}{M2M}{Machine to Machine}
\newacronym{mac}{MAC}{Medium Access Control}
\newacronym{mc}{MC}{Multi-Connectivity}
\newacronym{mcs}{MCS}{Modulation and Coding Scheme}
\newacronym{mec}{MEC}{Mobile Edge Cloud}
\newacronym{mi}{MI}{Mutual Information}
\newacronym{mimo}{MIMO}{Multiple-Input Multiple-Output}
\newacronym{mmwave}{mmWave}{millimeter wave}
\newacronym{mr}{MR}{Maximum Rate}
\newacronym{mss}{MSS}{Maximum Segment Size}
\newacronym{mtd}{MTD}{Machine-Type Device}
\newacronym{mtu}{MTU}{Maximum Transmission Unit}
\newacronym{nfv}{NFV}{Network Function Virtualization}
\newacronym{nlos}{NLOS}{Non Line of Sight}
\newacronym{nr}{NR}{New Radio}
\newacronym{ofdm}{OFDM}{Orthogonal Frequency Division Multiplexing}
\newacronym{pdcch}{PDCCH}{Physical Downlonk Control Channel}
\newacronym{pdcp}{PDCP}{Packet Data Convergence Protocol}
\newacronym{pdsch}{PDSCH}{Physical Downlink Shared Channel}
\newacronym{pdu}{PDU}{Packet Data Unit}
\newacronym{pf}{PF}{Proportional Fair}
\newacronym{pgw}{PGW}{Packet Gateway}
\newacronym{phy}{PHY}{Physical}
\newacronym{pbch}{PBCH}{Physical Broadcast Channel}
\newacronym[plural=\gls{mme}s,firstplural=Mobility Management Entities (MMEs)]{mme}{MME}{Mobility Management Entity}
\newacronym{prb}{PRB}{Physical Resource Block}
\newacronym{pss}{PSS}{Primary Synchronization Signal}
\newacronym{pucch}{PUCCH}{Physical Uplink Control Channel}
\newacronym{pusch}{PUSCH}{Physical Uplink Shared Channel}
\newacronym{rach}{RACH}{Random Access Channel}
\newacronym{ran}{RAN}{Radio Access Network}
\newacronym{red}{RED}{Random Early Detection}
\newacronym{rf}{RF}{Radio Frequency}
\newacronym{rlc}{RLC}{Radio Link Control}
\newacronym{rlf}{RLF}{Radio Link Failure}
\newacronym{rrc}{RRC}{Radio Resource Control}
\newacronym{rrm}{RRM}{Radio Resource Management}
\newacronym{rr}{RR}{Round Robin}
\newacronym{rs}{RS}{Remote Server}
\newacronym{rsrp}{RSRP}{Reference Signal Received Power}
\newacronym{rss}{RSS}{Received Signal Strength}
\newacronym{rtt}{RTT}{Round Trip Time}
\newacronym{rw}{RW}{Receive Window}
\newacronym{rx}{RX}{Receiver}
\newacronym{sa}{SA}{standalone}
\newacronym{sack}{SACK}{Selective Acknowledgment}
\newacronym{sap}{SAP}{Service Access Point}
\newacronym{sch}{SCH}{Secondary Cell Handover}
\newacronym{scoot}{SCOOT}{Split Cycle Offset Optimization Technique}
\newacronym{sdma}{SDMA}{Spatial Division Multiple Access}
\newacronym{sinr}{SINR}{Signal to Interference-plus-Noise Ratio}
\newacronym{sm}{SM}{Saturation Mode}
\newacronym{snr}{SNR}{Signal to Noise Ratio}
\newacronym{son}{SON}{Self-Organizing Network}
\newacronym{ss}{SS}{Synchronization Signal}
\newacronym{srs}{SRS}{Sounding Reference Signal}
\newacronym{sss}{SSS}{Secondary Synchronization Signal}
\newacronym{tb}{TB}{Transport Block}
\newacronym{tcp}{TCP}{Transmission Control Protocol}
\newacronym{tdd}{TDD}{Time Division Duplexing}
\newacronym{tdma}{TDMA}{Time Division Multiple Access}
\newacronym{tfl}{TfL}{Transport for London}
\newacronym{tm}{TM}{Transparent Mode}
\newacronym{trp}{TRP}{Transmitter Receiver Pair}
\newacronym{tti}{TTI}{Transmission Time Interval}
\newacronym{ttt}{TTT}{Time-to-Trigger}
\newacronym{tx}{TX}{Transmitter}
\newacronym{ue}{UE}{User Equipment}
\newacronym{ul}{UL}{Uplink}
\newacronym{uml}{UML}{Unified Modeling Language}
\newacronym{um}{UM}{Unacknowledged Mode}
\newacronym{utc}{UTC}{Urban Traffic Control}
\newacronym{vm}{VM}{Virtual Machine}
\newacronym{rsrq}{RSRQ}{Reference Signal Received Quality}
\newacronym{rssi}{RSSI}{Received Signal Strength Indicator}
\newacronym{crs}{CRS}{Cell Reference Signal}
\newacronym{comp}{CoMP}{Coordinated Multi-Point}
\newacronym{cran}{C-RAN}{Cloud \acrlong{ran}}
\newacronym{ca}{CA}{Carrier Aggregation}
\newacronym{cco}{CC}{Carrier Component}
\newacronym{nsa}{NSA}{Non Stand Alone}
\newacronym{embb}{eMBB}{Enhanced Mobility Broadband}
\newacronym{bsr}{BSR}{Buffer Status Report}
\newacronym{srb}{SRB}{Service Radio Bearer}
\newacronym{scm}{SCM}{Spatial Channel Model}
\newacronym{sctp}{SCTP}{Stream Control Transmission Protocol}
\newacronym{mptcp}{MPTCP}{Multi-path TCP}
\newacronym{ietf}{IETF}{Internet Engineering Task Force}
\newacronym{os}{OS}{Operating System}
\newacronym{tls}{TLS}{Transport Layer Security}
\newacronym{rfc}{RFC}{Request for Comments}
\newacronym{http}{HTTP}{HyperText Transfer Protocol}
\newacronym{nat}{NAT}{Network Address Translation}
\newacronym{api}{API}{Application Programming Interface}
\newacronym{rto}{RTO}{Retransmission Timeout}
\newacronym{psc}{PSC}{Public Safety Communication}
\newacronym{rpgm}{RPGM}{Reference Point Group Mobility}
\newacronym{ic}{IC}{Incident Command}
\newacronym{rsu}{RSU}{Road Side Unit}
\newacronym{uav}{UAV}{Unmanned Aerial Vehicle}
\newacronym{iab}{IAB}{Integrated Access and Backhaul}
\newacronym{psd}{PSD}{Power Spectral Density}
\newacronym{mpc}{MPC}{Multi Path Component}
\newacronym{rt}{RT}{Ray Tracer}
\newacronym{aoa}{AoA}{Angle of Arrival}
\newacronym{aod}{AoD}{Angle of Departure}
\newacronym{inr}{INR}{Interference to Noise Ratio}
\newacronym{qd}{QD}{Quasi Deterministic}
\newacronym{wlan}{WLAN}{Wireless Local Area Network}
\newacronym{cad}{CAD}{Computer-aided Design}
\newacronym{ap}{AP}{Access Point}
\newacronym{sta}{STA}{Station}
\newacronym{nrmse}{NRMSE}{Normalized Root Mean Square Error}
\newacronym{ut}{UT}{User Terminal}
\newacronym{bs}{BS}{Base Station}

\copyrightyear{2022}
\acmYear{2022}
\setcopyright{none}
\acmConference[WNS3 2022]{2022 Workshop on ns-3}{June 22--23, 2022}{Virtual Event, USA}
\acmBooktitle{2022 Workshop on ns-3 (WNS3 2022), June 22--23, Virtual Event}
\acmPrice{xxx}
\acmDOI{xxx}
\acmISBN{xxx}

\begin{document}
\pagestyle{empty}
\title{ns-3 and 5G-LENA Extensions to Support Dual-Polarized MIMO}

\author{Biljana Bojovic, Zoraze Ali, Sandra Lagen}
\affiliation{\institution{Centre Tecnol\`ogic de Telecomunicacions de Catalunya (CTTC/CERCA), Barcelona, Spain}
  \postcode{08013}
}
\email{{biljana.bojovic, zoraze.ali, sandra.lagen}@cttc.es}

\pagestyle{empty}

\begin{abstract}
MIMO spatial multiplexing is an essential feature to increase the communication data rates in current and future cellular systems. 
Currently, the ns-3 {lte} module leverages an abstraction model for 2x2 MIMO with spatial multiplexing of two streams; while {mmwave} and {nr} modules were lacking the spatial multiplexing option until this work, since the ns-3 models were not supporting the usage of multiple antennas for spatial multiplexing and an abstraction model such as the one used in the {lte} module is not suitable for the mmWave frequencies. In this paper, we propose, implement and evaluate models for ns-3 and the {nr} module to enable Dual-Polarized MIMO (DP-MIMO).
The proposed extension for the ns-3 supports multiple antennas for DP-MIMO with spatial multiplexing of two streams and can be used by any ns-3 module that is compatible with the ns-3 antenna array-based models, such as {nr} and {mmwave} modules. We leverage this ns-3 extension to model DP-MIMO by exploiting dual-polarized antennas and their orthogonality under line-of-sight conditions, as it happens at high-frequency bands, to send the two data streams. The proposed model does not rely on abstraction, as the MIMO model in the ns-3 {lte} module, and can thus model more realistically the propagation differences of the two streams, correlation, inter-stream interference, and allows design and evaluation of the rank adaptation algorithms. Additionally, we propose and evaluate an adaptive rank adaptation scheme and compare it with a fixed scheme. The developed DP-MIMO spatial multiplexing models for the ns-3 simulator and the {nr} module are openly available. \end{abstract}


\maketitle

\section{Introduction}\label{sec:intro}
\glsresetall

\begin{picture}(0,0)(0,-580)
\put(0,0){
\put(0,0){\small This paper has been submitted to WNS3 2022. Copyright may be transferred without notice.}}
\end{picture}

The use of multiple antennas in communication systems (a.k.a. \gls{mimo} systems) has attracted a lot of attention in the recent decades. MIMO permits~\cite{842121}:
\begin{itemize}
    \item increasing the data rate by sending multiple data streams simultaneously (known as \textit{spatial multiplexing}), thanks to the use of various \gls{rf} chains,
    \item increasing the robustness of the data transmission by sending replicated data (known as \textit{transmit diversity}), or 
    \item increasing the \gls{sinr} by providing array gain (known as \textit{beamforming}), thanks to properly designing the antenna weights.
\end{itemize}
At high carrier frequencies within the \gls{mmwave} region, beamforming is particularly essential to combat the high pathloss propagation losses and blocking effects~\cite{pi:11}. In case of beamforming, a single spatial stream is sent per receiver and the multiple antennas are used to concentrate the radiated power towards the receiver's location, thus improving the received \gls{sinr} and the probability of error at the target receiver, as well as reducing the generated interference towards other spatial locations. 
To further increase the user data rates, MIMO spatial multiplexing is required, to allow sending multiple data streams per user. However, \gls{mmwave} systems may use only a small number of radio-frequency (RF) chains (and so, low number of spatial streams) due to the high cost and power utilization of RF chains working at high carrier frequencies. 

In the literature, antenna polarization has been proposed as an attractive strategy to realize MIMO spatial multiplexing~\cite{5638598}. In particular, advanced polarization techniques such as dual-polarized antennas (leading to Dual-Polarized MIMO (DP-MIMO) systems) have been shown to improve the spectral efficiency of multiple antenna systems as compared to classical MIMO systems, for low levels of cross polarization discrimination~\cite{6206454}. Dual-polarized antennas are indeed a practical and effective way to enable MIMO spatial multiplexing at mmWave frequencies, because only two RF chains are needed, which is suitable for real implementations in those bands, and because propagation is characterized by low levels of cross polarization discrimination. Basically, in DP-MIMO at mmWave bands, two streams can be simultaneously sent across the two orthogonal polarizations (i.e., vertical (0) and horizontal ($+$90) polarizations, or $+$45 and $-45$ polarizations), as considered by 3GPP~\cite{TR38901}.

The ns-3 {mmwave} module for 5G NR in \gls{mmwave} bands~\cite{mezzavilla2017end}, as well as the ns-3 {nr} (5G-LENA) module for 5G NR in sub 6 and \gls{mmwave} bands~\cite{PATRICIELLO2019101933} support various ideal beamforming methods. Recently, we developed realistic beamforming methods~\cite{bil21ns3}, for which real resources are employed to perform beam management, considering a non-ideal channel estimation. However, both modules have been lacking up to now MIMO spatial multiplexing feature. 
In ns-3 {lte} (LENA) module there is an abstraction model to simulate 2$\times$2 MIMO, which models up to two spatial streams per user. However, such a model does not account for propagation differences among different streams, and assumes no correlation between antennas, which is a non-realistic assumption for \gls{mmwave} bands.

In this paper, relying on dual-polarized antennas, we propose, implement, and evaluate a 2-stream MIMO model for NR. The model is specifically suited for high frequency bands (i.e., \gls{mmwave}), because it assumes the use of only two RF chains at gNB and UE nodes. Even so, it can be also applied to sub 6 GHz bands without modifications, just with the inherent limitation of two spatial streams per user.

The rest of the paper is organized as follows. Sec.~\ref{sec:mimo_ns3} reviews MIMO antecedent implementations in ns-3. Sec.~\ref{sec:mimo} describes the adopted MIMO model for 5G NR at high frequency bands, which combines spatial multiplexing and beamforming. Sec.~\ref{sec:ns3} presents the ns-3 implementation details of the proposed MIMO framework, including the changes in the {antenna} and the {spectrum} module, and the PHY and MAC layers of the {nr} module. Finally, Sec.~\ref{sec:examples} details the simulation example, and Sec.~\ref{sec:results} presents the simulation results for validation.

\section{MIMO antecedents in ns-3}
\label{sec:mimo_ns3}
The ns-3 {lte} module includes abstracted MIMO models for spatial multiplexing and transmit diversity~\cite{lena}. The model is based on a statistical gain of several MIMO solutions with respect to the SISO (Single-Input Single-Output) one~\cite{1203169}, assuming no correlation between the antennas. It supports various MIMO schemes, namely, SISO, MIMO-Alamouti, MIMO-MMSE, MIMO-OSIC-MMSE and MIMO-ZF schemes. In the case of MIMO spatial multiplexing, the PHY abstraction model is limited to 2 streams per UE. At the MAC layer, multiple MAC transport blocks are created, one per stream, and this is considered in the scheduler as well.
The no correlated antennas assumption does not hold for mmWave bands, in which indeed the MIMO channel can be rank deficient, with correlated antenna elements. So, these models are not applicable there. 

The ns-3 {mmwave} module~\cite{mezzavilla2017end} introduced various MIMO beamforming methods for mmWave bands. Namely, the long-term covariance matrix method, the beam search method and the line-of-sight (LOS) path method. In the long-term covariance matrix method, beamforming vectors are computed from the maximal eigenvectors of the channel covariance matrix. In the beam search method, it is assumed that there is a discrete number of beams from a pre-designed codebook, from which the beam that provides the largest SINR (considering the beam and the channel matrix) is selected for each gNB-UE link. In the LOS path method the DoA is assumed to be perfectly known and used to calculate the beamforming vectors to steer the beam towards such direction. 

In the ns-3 {nr} module~\cite{PATRICIELLO2019101933}, we had available MIMO beamforming methods, as inherited from the {mmwave} module, i.e., the beam search method and the LOS path method. Later on, we introduced the representation for a quasi-omnidirectional beamforming, and multiple combinations with the previous methods (e.g., the gNB using beam search method and the UE using quasi-omni reception). Recently, we went a step over the ideal beamforming methods, and implemented realistic beamforming methods using SRS-based channel estimates~\cite{bil21ns3}. That is, we used SRS (transmitted in specific time/frequency resources) to estimate the channel matrix, and then determine the beamforming vectors at the gNB based on the SRS-based channel estimate. 

However, all the methods available in ns-3 NR modules ({mmwave} and {nr}) ~\cite{mezzavilla2017end,PATRICIELLO2019101933,bil21ns3} up to date, were focusing on MIMO beamforming and they were missing the support of MIMO spatial multiplexing. With the contribution of this paper, we start to fill this gap.


\section{DP-MIMO modeling for mmWave bands}
\label{sec:mimo}
3GPP considers gNB/UE nodes equipped with dual-polarized uniform planar antenna arrays~\cite{TR38901}. In such systems, each array
of antennas is divided evenly into two subarray groups with different polarizations (e.g., one subarray of vertically
polarized antennas and the other subarray of horizontally polarized antennas). The gNB/UE have two RF chains, one for each polarization. In addition, assume that each antenna subarray (or polarization) is precoded with its own beamforming vector, to overcome mmWave propagation limits.
This way, the MIMO model adopted in 5G-LENA, can combine spatial multiplexing (with up to two spatial streams per user) and beamforming (applied to each spatial streams separately). 

\begin{figure}
    \centering
    \includegraphics[width=0.45\textwidth]{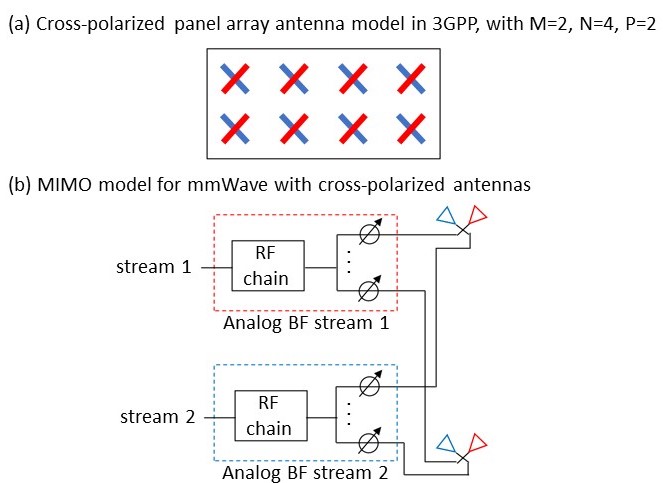}
    \vspace{-0.3cm}
    \caption{(a) dual-polarized antenna model in 3GPP and (b) DP-MIMO model exploiting orthogonal polarizations}
    \label{fig:mimo}
    \vspace{-0.3cm}
\end{figure}

Figure~\ref{fig:mimo}.(a) illustrates the cross-polarized antenna model considered in 3GPP~\cite{TR38901}. The 3GPP model considers that in each panel, antenna elements are placed in the vertical and horizontal direction, where $N$ is the number of columns and $M$ is the number of antenna elements with the same polarization in each column. The antenna elements are uniformly spaced in the horizontal direction with a certain spacing and in the vertical direction with a certain spacing. The antenna panel can be either single polarized ($P=1$) or dual-polarized ($P=2$). In case of $P=1$, all the antenna elements are vertically polarized (with polarization slant angle of 0). In case of $P=2$, a subset of the antenna elements are polarized with polarization slant angle of $+45^{\circ}$ (red elements in the figure), and the rest with $-45^{\circ}$ (blue elements in the figure). Figure~\ref{fig:mimo}.(a) shows an example for $M=2$, $N=4$, and $P=2$.

Figure~\ref{fig:mimo}.(b) shows how the dual-polarized antennas can be used for a transceiver design that supports MIMO spatial multiplexing. In this case, we have two RF chains, associated with two analog beamforming vectors, that connect to the antenna elements of a given polarization. Each spatial stream is associated to one of the polarizations. In the figure, stream 1 is sent/received through/by the $+45^{\circ}$ polarized antennas, and stream 2 is sent/received through/by the $-45^{\circ}$ polarized antennas.

\begin{figure}
    \centering
    \includegraphics[width=0.38\textwidth]{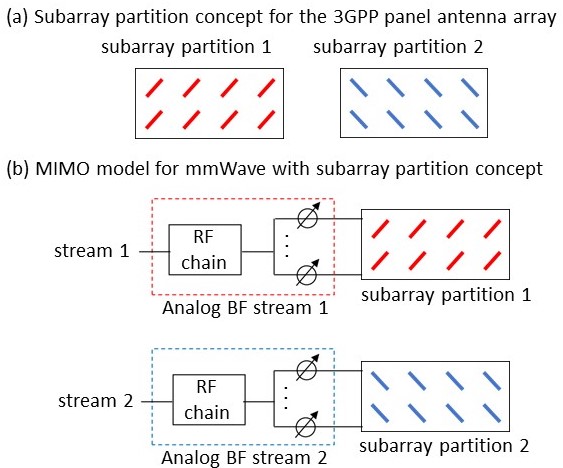}
    \caption{Subarray partition concept, applied to (a) dual-polarized antenna model in 3GPP and (b) DP-MIMO model }
    \label{fig:mimo2}
\end{figure}

Finally, Figure~\ref{fig:mimo2} shows the subarray partition concept, when applied to the 3GPP antenna array and the full MIMO model architecture for mmWave. Basically, antennas with same polarization are mapped to the same antenna subpartition, as shown in Figure~\ref{fig:mimo2}.(a). The two subpartitions are collocated in space (see Figure~\ref{fig:mimo}), just here are separated in a logical manner. Our implementation in ns-3 follows this same concept. Based on the logical partitions, it is easy to associate spatial streams to subpartitions, as graphically illustrated in Figure~\ref{fig:mimo2}.(b).

\section{ns-3 and {nr} implementations for MIMO}
\label{sec:ns3}
This section provides the MIMO implementation details, including the ns-3 changes to support multiple antenna subarrays (in {antenna} and {spectrum} modules), and changes to the {nr} module to support MIMO spatial multiplexing (mostly in {PHY} and {MAC} layers).

\subsection{ns-3 antenna}
\label{sec:ant}
We extended the {UniformPlanarArray} class of {antenna} module, to consider the polarization slant angle. In particular, the attribute {PolSlantAngle} is the polarization slant angle that applies to all elements of a subarray partition. This value has to be configured separately for each subarray partition, e.g., with $+45$ for the first subarray partition and with $-45$ for the second subarray partition. This is already included in ns-3-dev.

\subsection{ns-3 spectrum}
\label{sec:spec}
The major changes to support the MIMO architecture for mmWave are introduced in the {spectrum} module. These changes aim at extending the current spectrum framework to support multiple array subpartitions per gNB/UE devices.
To support such collocated antenna arrays (subpartitions) per device we needed to extend the {ThreeGppChannelModel} to be able to distinguish the channel parameters that are common for all the channels among the same pair of the transmit/receive (TX/RX) nodes and those that are specific for the TX/RX antenna subpartition array pair. 
For example, if the TX and RX nodes have multiple collocated antenna arrays, there will be multiple channel matrices among the same pair of nodes for the different pairs of the TX and RX antenna subarrays.
However, these channel matrices that are among the same pair of nodes have common channel parameters, i.e., they share the same channel condition, cluster powers, cluster delays, AoD, AoA, ZoD, ZoA, K\_factor, delay spread, etc.~\cite{TR38901}. Hence, these parameters should not be regenerated for each pair of antenna subarrays among the same pair of the TX and RX nodes.
Additionally, each pair of the TX and RX antenna subarrays has a specific channel matrix and fading, which depends on the actual antenna element positions and field patterns of each pair of antenna array subpartitions. 

\begin{figure*}
\centering
\subfloat[]
{
\includegraphics[width=0.28\textwidth]{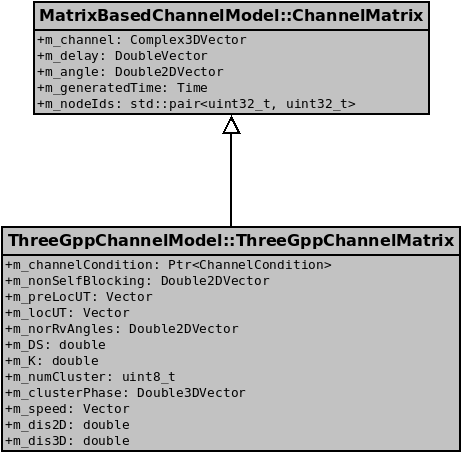}
\label{fig:Channel matrix structures before the split }
}
\subfloat[] { 
\includegraphics[width=0.65\textwidth]{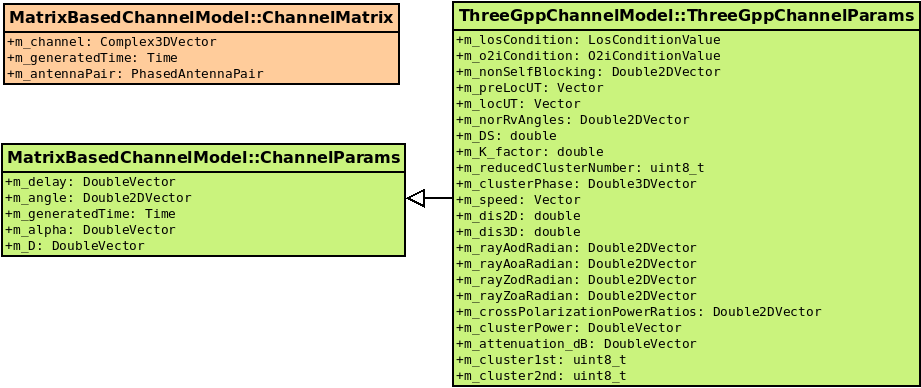}
\label{fig:Channel matrix structures after the split }
}
\caption{Splitting the channel matrix and the channel parameters into the two structures to support MIMO}
\label{fig:foo}
\end{figure*}

For that reason, we split the function {GetNewChannel} into GetNewChannelParams and GetNewChannelMatrix, which update the respective parameters. Accordingly, the channel parameters are saved into two separate structures ({ChannelParams} - per node pair and {ChannelMatrix} - per phased antenna array pair).
Figure \ref{fig:foo} illustrates the main channel structures (a) before and (b) after the split. 
Grey color indicates the structures that existed before the split. Orange color denotes the existing structure but updated. Green color shows the two new structures, one generic, implemented in the base class {MatrixBasedChannelModel}, and its 3GPP specialization implemented in {ThreeGppChannelModel} class. Notice that, the previously existing structure called {ThreeGppChannelMatrix} does not exist anymore, instead, most of its fields have been moved to {ThreeGppChannelParams} structure.

Additionally, to allow multiple antenna arrays per device (and per {SpectrumChannel} instance), we extended the {spectrum} module to support passing the TX and RX antenna arrays objects to the spectrum propagation loss models that require pointers to TX and RX antenna array objects in order to calculate RX PSD, like e.g. ThreeGppSpectrumPropagationLossModel, which needs the beamforming vectors to calculate RX PSD. 
To achieve this, we created a new type of spectrum propagation loss model interface called {PhasedArraySpectrumPropagationLossModel}, in which its function {CalcRxPowerSpectralDensity} has as input also the pointers to {PhasedArrayModel} objects of the TX and RX. {ThreeGppSpectrumPropagationLossModel} class now inherits this new interface.

\begin{figure*}
    \centering
    \includegraphics[width=0.95\textwidth]{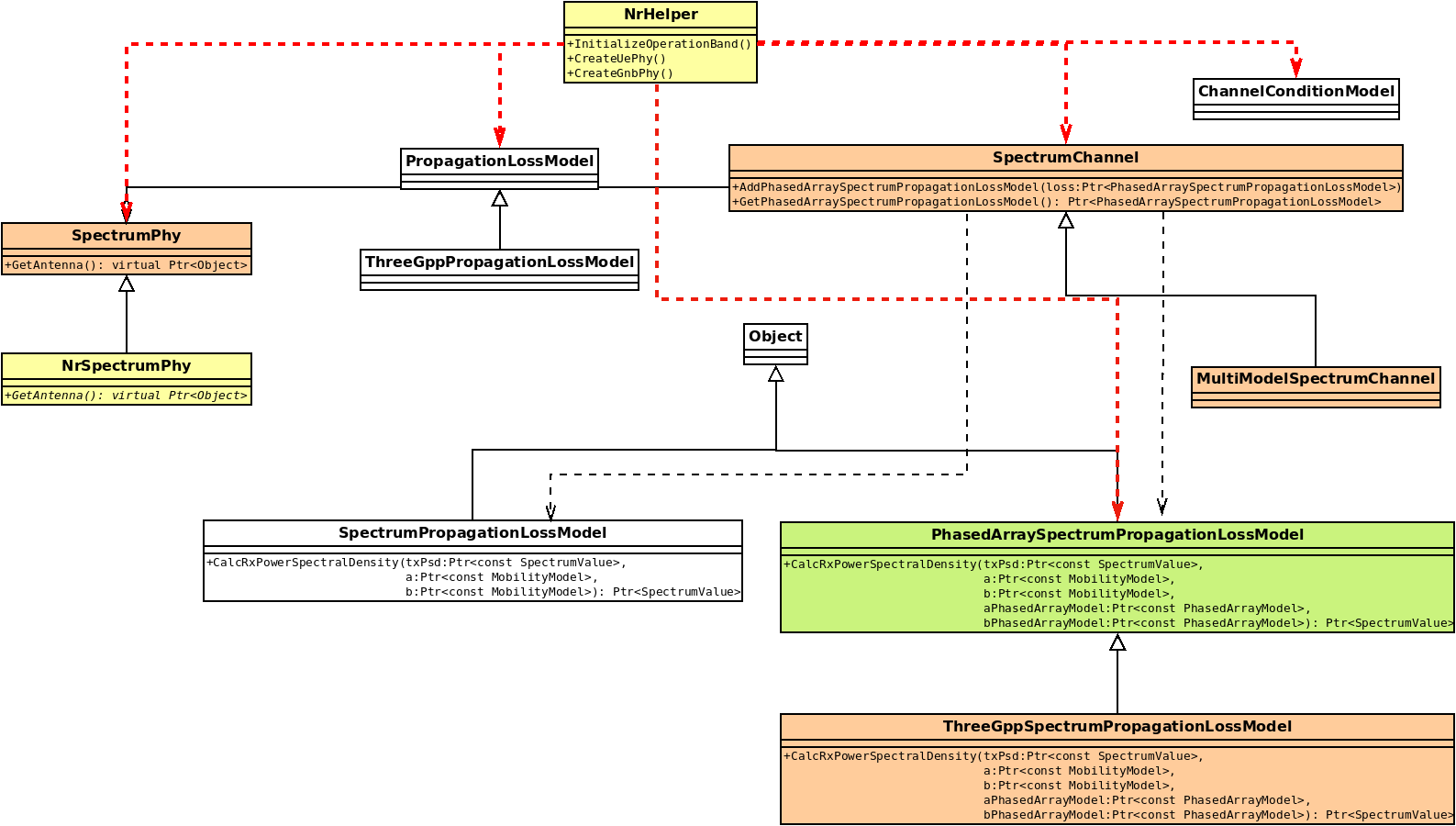}
    \caption{Changes in the spectrum module to support MIMO}
    \label{fig:mimo-spectrum}
\end{figure*}

By adding this very simple design change into the {spectrum} module, we removed a hard limitation of ThreeGppSpectrumPropagationLossModel which was to support a maximum of 1 antenna array instance per ns-3 device (per {SpectrumChannel} instance).
According to the new {spectrum} design, all devices that use antenna arrays and related spectrum models, should configure {PhasedAntennaArray} antenna per {SpectrumPhy} instance and attach the {SpectrumPhy} instance to the channel of interest. 
In this way, devices can have several antenna arrays transmitting and receiving on the same {SpectrumChannel}. Figure~\ref{fig:mimo-spectrum} illustrates these extensions.

In the MultiModelSpectrumChannel::StartTx function we added a condition that checks whether the TX/RX {SpectrumPhy} instances belong to the different nodes. This is needed to avoid pathloss calculations among the antenna arrays of the same node, because there are no models yet in ns-3 for channel modeling among the antenna arrays of the same node.

The proposed changes to {spectrum} module generalize the previous functionality, i.e., multiple antenna arrays are permitted per device and {SpectrumChannel}, while the behaviour when a single antenna array is configured remains the same.
However, due to the changes related to {ThreeGppSpectrumPropagationLossModel} API, the modules that use this API need to override the function {SpectrumPhy::GetAnntenna} to return the {PhasedAntennaArray} corresponding to that {SpectrumPhy} instance. 
Figure~\ref{fig:mimo-spectrum} shows that, e.g., {NrSpectrumPhy} inherits {SpectrumPhy} and implements this function. According to the proposed API, {GetRxAntenna} function returns a pointer to {Object} so that it can support two different types of antenna models in ns-3, i.e., those based on {AntennaModel} API as well as {PhasedAntennaArray} API.

The proposed {spectrum} module's extensions are valuable for ns-3 because by allowing multiple antenna arrays per device on the same {SpectrumChannel}, modules using this propagation model can be extended to support MIMO through dual-polarized antennas, i.e., as we explain in this paper on the example of the {nr} module. Also, this model can be used for MU-MIMO, by considering multiple subarrays to transmit streams towards different users through the array subpartition concept. Furthermore, it could enable the hybrid centralized RAN architecture, with a gNB node controlling multiple radio units.

\subsection{ns-3 nr PHY}
\label{sec:phy}
In the following, we describe the {nr} PHY layer extensions to support MIMO spatial multiplexing.
\subsubsection{RI computation and rank adaptation algorithm}
In real systems, a gNB capable of performing MIMO spatial multiplexing, e.g., in the \gls{dl}, can use more than one stream to transmit to those UEs that support MIMO spatial multiplexing. However, gNB's decision to use multiple streams depends on the channel Rank Indicator (RI) reported by a UE. We implemented two schemes for computing the RI for DL MIMO transmissions: fixed and adaptive.
\paragraph*{Fixed RI scheme}
The fixed RI scheme uses a fixed RI configured by the user of the simulator. To enable this scheme, we added {UseFixedRi} and {FixedRankIndicator} attributes, in the {NrUePhy} class. For example, to use a fixed RI of 1 for a particular UE or all the UEs throughout the simulation, one can configure $UseFixedRi=true$ and $FixedRankIndicator=1$. This configuration entails that irrespective of the UE's support for MIMO (i.e., it has two subarrays), the gNB throughout the simulation will use only one stream (i.e., the first subarray) to transmit to the UEs reporting fixed RI of 1.
\paragraph*{Adaptive RI scheme}
The adaptive RI scheme uses an adaptive algorithm based on two SINR thresholds to compute an RI value~\cite{6364098}. We have considered two SINR thresholds to differ the case of transitioning from 2 to 1 streams and that of going from 1 to 2, because of the power distribution change. It is worth mentioning, other simulators that involve a detailed implementation of physical layer procedures, e.g., link-layer level simulators, perform complex operations, such as Singular Value Decomposition (SVD) on a channel matrix to compute the Rank of the channel. However, these matrix operations cannot be performed in ns-3 without the support of external libraries (if any). In addition, it could increase the simulation time that grows exponentially with the number of channel objects in a simulation. Therefore, we propose to use SINR to achieve a simple and less computationally complex RI adaptation algorithm(s). This algorithm compares the SINR of the data channel (i.e., PDSCH) of each active stream for which UE has received the data with two preconfigured thresholds to determine the RI value (1 or 2).  To enable this adaptive RI algorithm, one must set the {NrUePhy} attribute, {UseFixedRi}, to false. Moreover, these two SINR thresholds, namely, {RiSinrThreshold1} and {RiSinrThreshold2}, are attributes of the {NrUePhy} class and can be configured as required. In detail, a UE that has the SINR of one active stream compares it with {RiSinrThreshold1}. If it is above this threshold, the UE has excellent propagation conditions, and it can report an RI value of 2 (i.e., switching from one to two streams).
On the other hand, if a UE is already receiving data using two streams, it compares the SINR of each stream with {RiSinrThreshold2}. If the SINR of both the streams is above this threshold, it keeps reporting an RI value of 2; otherwise, it uses an RI value of 1 (i.e., switching from two streams to one). We note that when a UE switches from two streams to one, besides reporting an RI value of 1, it also reports the CQI of both the streams to the gNB. This allowed the extension of the scheduler to choose the stream with better CQI for future allocations. 

\subsubsection{CQI and RI reporting}
{NrUePhy} already supports reporting of a single wide-band CQI value to a gNB under SISO communication. It uses a C$++$ structure {DlCqiInfo} that mimics the information element {cqiListElement} of FemtoForum MAC scheduler interface specifications~\cite{ff_mac_sched}. However, with the introduction of MIMO, {NrUePhy} should communicate the CQI of each MIMO stream for which it has received the data. Therefore, we replaced the member variable of this structure used to store one CQI value with a C$++$ vector capable of holding the CQI of multiple streams. In particular, there is a direct mapping between a stream index and the indexes of this vector. For example, the CQI of the first stream is stored at the index 0 of this vector because the streams are indexed from zero in our simulator. Finally, the {DlCqiInfo} structure includes a variable {m\_ri} specifically to report the RI value to the gNB.

\begin{figure}
    \centering
    \includegraphics[width=0.45\textwidth]{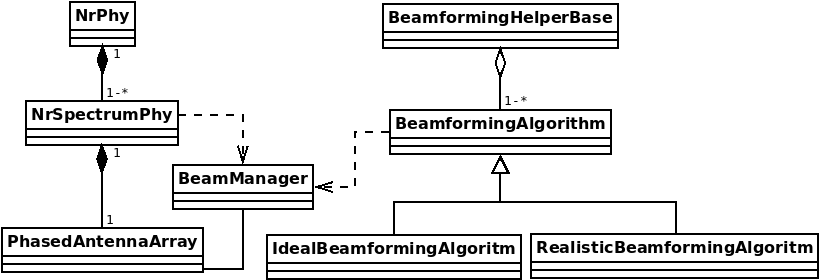}
    \caption{Changes in NR PHY to support MIMO: multiple antenna arrays per PHY and the beamforming management}
    \label{fig:mimo-nr-phy}
\end{figure}

\subsubsection{nr PHY TX/RX through multiple streams}
To support multiple streams, the {NrPhy} class is extended to aggregate multiple {NrSpectrumPhy} instances, and there is one PhasedAntennaArray per each {NrSpectrumPhy} instance.
Each instance of {PhasedAntennaArray} class models a subpartition of the antenna array, and each subpartition is used for a specific stream. 
This is illustrated in Figure~\ref{fig:mimo-nr-phy}.
To achieve the proposed MIMO spatial multiplexing configuration, we install two {NrSpectrumPhy} instances per {NrGnbPhy} and {NrUePhy}, and then we configure the two {UniformPlannarArray} instances, two antenna array subpartitions, 
belonging to the same {NrGnbPhy} or {NrUePhy} to be cross polarized.
For this, the extension of the ns-3 {UniformPlannarAray} is used, described in Sec.~\ref{sec:ant}. Even if there are 2 subpartitions of cross polarized antenna arrays at both gNB and UE, with the current implementation, both streams will be used for the DATA/CTRL transmission only in the DL. The \gls{ul} is configured to use a single stream for CTRL and DATA, while SRS is transmitted in both streams in UL to allow the realistic beamforming functionality.

 \subsubsection{Beamforming per antenna subpartition}
To support the beamforming per subpartition of antenna array, we changed the NR PHY model to have a {BeamManager} per {NrSpectrumPhy}, instead of having a single {BeamManager} per {NrUePhy} or {NrGnbPhy}. This was necessary because one {BeamManager} instance supports a single beamforming vector configuration (hence single subpartition). Additionally, the whole NR module's beamforming framework, including ideal and realistic algorithms, and corresponding helpers, we extended to support multiple antenna arrays or subpartitions per t{NrUePhy} or {NrGnbPhy} instance, so that the beamforming is performed independently per each subpartition. Beamforming functionality in {nr} module assumes that the beamforming should be performed among the subpartitions with the same index, i.e., the beamforming is performed among the 1st subarray of the gNB and the 1st subarray of the UE, and among the 2nd subarray of the gNB and 2nd subarray of the UE. If a gNB supports 2 streams, and a UE only 1 stream, the beamforming is done among the 1st subarray of the gNB and the UE's subarray. Also, since SRS is reported per each stream, NR module's realistic beamforming is updated to be performed per each stream. Figure \ref{fig:mimo-nr-phy} illustrates the {nr} module beamforming model support for DP-MIMO.

 \subsubsection{HARQ and SINR reporting for multiple streams}
The {nr} PHY model,  
including {NrSpectrumPhy} and {NrUePhy}, is extended to support Hybrid Automatic Repeat Request (HARQ) and SINR reporting for multiple streams.
The {nr} PHY functions are extended to include the index of the stream when e.g., HARQ is reported, and the index is used to merge the feedbacks from different streams into the new HARQ structure. 
Similarly, SINR reporting is updated to support multiple streams, i.e., when {NrSpectrumPhy} notifies {NrUePhy} about SINR, it indicates to which streams this SINR corresponds. Previously, some of the interference traces were hooked to {NrUePhy}, because there was only 1 {NrSpectrumPhy} per {NrUePhy}. But, since now we may have multiple {NrSpectrumPhy} instances per {NrUePhy}, we connected the SINR traces coming from {NrInterference} and {ChunkProcessors} directly to {NrSpectrumPhy}, and then NrSpectrumPhy passes these traces to {NrUePhy} by indicating also the stream to which corresponds the trace.  

 \subsubsection{TX power per stream}
When setting the gNB transmit power for each transmission, it is taken into account that the configured power is the total power of all active streams of that {NrGnbPhy}. So, the transmit power per stream is being dynamically determined
before each transmission and depends on the number of the active streams, i.e., the total transmit power is divided by the number of the active streams.
 
\subsubsection{Inter-stream interference}
Another important aspect that we studied is how to handle inter-stream interference. We implemented two different ways to configure and tune inter-stream interference. First, we implemented a new channel model called ThreeGppChannelModelParam, based on {ThreeGppChannelModel}, with which we can parametrize the inter-stream interference correlation, based on the 3GPP cross-polarization correlation parameter. Secondly, we added in {NrSpectrumPhy} an attribute called InterStreamInterferenceRatio which can tune the level of the inter-stream interference, and it depends on the assumed receiver's capability to handle this interference. The value of this parameter ranges from 0 to 1. 0 value means that there is no inter-stream interference, and 1 that the full inter-stream interference is taken into account, i.e., the RX PSD of the cross-polarized signal is considered when SINR is calculated. 
For example, we observed when assuming full inter-stream interference ({InterStreamInterferenceRatio = 1}) that in scenarios of no cross-polarization correlation (i.e., ThreeGppChannelModelParam configured in a such way that there is no correlation among the streams of different polarizations), the inter-stream interference goes to 0, while in other scenarios, interference is present. 

\subsubsection{Support for OFDMA scheduling}
 In the {nr} module the OFDMA scheduling has the constraint that at the same variable TTI can be served simultaneously the UEs for which a gNB uses the same beam. To be able to distinguish among available beams, the {nr} module was relying on the concept of {BeamId}, which uniquely identifies a beam~\cite{NatlieMac}. 
 In order to support the OFDMA scheduling with MIMO spatial multiplexing while avoiding invasive changes of the OFDMA schedulers code, we created a {BeamConfId} structure based on {BeamId}, which identifies uniquely the pair of beams (one for each stream). In this way, only the UEs for which a gNB uses exactly the same "beam configuration" ({BeamConfId}), can be scheduled at the same time in the OFDMA manner. This implementation takes into account that some UEs can, in general, support a variable number of streams (1 or 2). 
Most of the {nr} extensions explained in this paper could support any number of streams, except the code for the OFDMA scheduling which depends on {BeamConfId} and thus supports up to two streams. Once in the future when this limitation is removed, the {nr} extensions for MIMO proposed in this paper will support any number of streams. 

\subsection{ns-3 nr MAC}
\label{sec:mac}
One of our main objectives while designing an {nr} module's support for MIMO spatial multiplexing was to keep intact the original SISO functionality. As a result, the proposed extension does not affect the core functionality of the MAC layer, nor the existing schedulers (e.g., RR, PF, and MR). To support the MAC scheduling of two transport blocks (one for each stream), we extended classes responsible for the core MAC functionalities, e.g., CQI management, DL Control Information (DCI) creation, HARQ feedback processing, etc. In the following, we explain these extensions.

\subsubsection{CQI management}
In the {nr} module, the class NrMacSchedulerCQIManagement is the final destination for DL and UL CQI reports. One of the main purposes of this class is to compute the DL and UL \gls{mcs} based on their respective CQIs. Since the MIMO support is introduced only for DL, the function {DlWBCQIReported} has been updated to read the new {DlCqiInfo} structure. Based on the CQI of each stream, it computes their MCS that is later used for deducing a TB size in NrMacSchedulerUeInfo::UpdateDlMetric.

\begin{figure} [!t]
    \centering
    \includegraphics[width=0.45\textwidth]{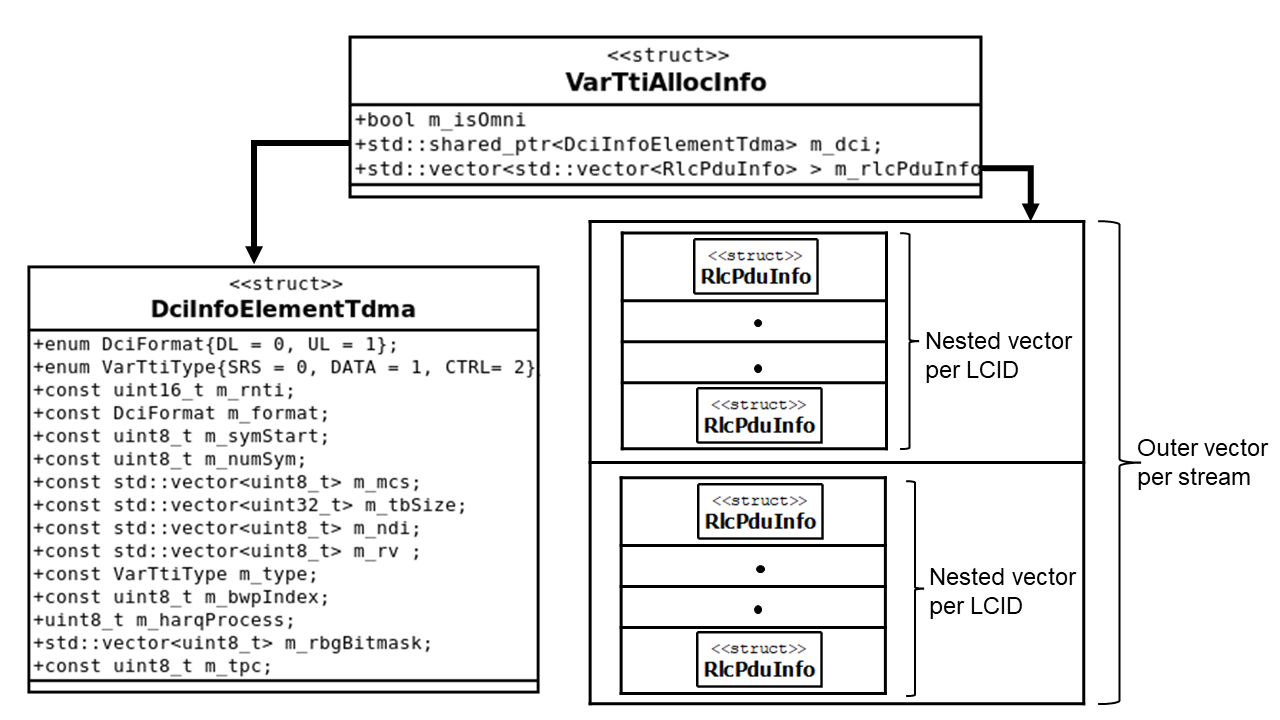}
    \caption{Updated VarTtiAllocInfo struct to support MIMO spatial multiplexing in 5G-LENA}
    \label{fig:dci}
\end{figure}

\subsubsection{DCI creation}
The NR module supports dynamic scheduled-based access both for DL and UL~\cite{NatlieMac}. 
A gNB is responsible for conveying the scheduling information (i.e., DL or UL) using a DCI to its UEs. A C++ struct, {VarTtiAllocInfo}, is currently used to communicate all the scheduling-related information needed by a gNB and a UE to orchestrate their transmissions and receptions. {VarTtiAllocInfo} structure contains: 1) {DciInfoElementTdma} structure, which represents the DCI, and 2) {RlcPduInfo}, which carries the number of bytes assigned to each active logical channels. Both structures were supporting originally the scheduling information for only one stream or TB. Therefore, to support multiple streams, we extended both, as shown in Fig.~\ref{fig:dci}. In {DciInfoElementTdma}, we extended the TB-related information variables, such as MCS, New Data Indicator (NDI), Redundancy Version (RV), and the TB size to be per stream or TB. We use a C++ vector for holding the TB-related information for each stream to allow that scheduler can simultaneously schedule the TBs of multiple MIMO streams. Similarly, to maintain the {RlcPduInfo} per MIMO stream, we wrapped the existing vector of {RlcPduInfo} into another vector per stream. The indexes of these vectors are analogous to the stream indexes at the {nr} PHY layer.

\subsubsection{MAC scheduling}

Thanks to the PHY layer extensions for MIMO, the CQI is reported per each stream. Hence, the scheduling of the new data is independent among the two streams and is performed by using the MCS of a specific stream. Also, the TB size may differ between the two streams.
This is a much more realistic behaviour than the one assumed in the current {lte}'s module abstraction model for MIMO. 
Also, if the CQIs are different between the two streams it may happen that a UE would be able to decode only one TB, which means that the NDI and RV information would not be the same for the two streams. Hence, the subsequent retransmissions (if any) are also independent. 

The number of streams to be used by the MAC scheduler for a specific UE depends on the RI value reported by the UE in a DL CQI report. In case that a UE reports an {RI = 1} (e.g., due to bad channel conditions), the proposed {nr} rank adaptation model is robust enough to choose a stream with a better CQI to schedule the upcoming transmission(s).
Notice that there are two situations in which {RI} value is not considered.
\begin{enumerate}
\item At the beginning of a simulation when the scheduler uses only one stream that belongs to the first subarray, even for the UE(s) that can support MIMO. This design choice is motivated by the two facts:
\begin{enumerate}
\item The current RRC model does not support UeCapabilityReport through which UE communicates its capability to support MIMO to a gNB, as in the standard.
\item In the {nr} module, CQI is computed using the data channel. Hence, a gNB should schedule at least one TB to receive an RI value reported in the CQI report, based on which it can decide whether to use one or two streams for the subsequent transmission(s). 
\end{enumerate}
\item During the retransmission phase. Specifically, the scheduler keeps rescheduling the TB of each stream independently until a UE can decode both streams or the maximum number of retransmissions (i.e., RV = 3) has been reached.

\end{enumerate}

\subsubsection{HARQ feedback processing}
The scheduler in the NR module is responsible for embedding a unique HARQ process id in every DCI it creates, and this functionality remains the same in the MIMO extension. Specifically, the simulator schedules multiple streams using a single DCI; thus, the same HARQ process id is used for all of them. It is worth mentioning that this implementation choice is aligned with the 3GPP standard. In particular, in the DCI format 1-1, which is a commonly used DCI for DL in NR, unlike TB-related information, the HARQ process id field is not repeated for each stream while using MIMO spatial multiplexing~\cite{DAHLMAN20181}. Moreover, similar to a real network, a UE in our simulator reports HARQ feedback (i.e., ACK or NACK) for each stream since it decodes them independently. To do that, we modified the C++ struct {DlHarqInfo}, which is used to report the HARQ feedback in downlink following the Femto forum MAC API~\cite{ff_mac_sched}. Specifically, this modification replaces the simple C++ enumeration (that represents ACK or NACK) with a vector of such enumeration so that a UE can report the HARQ feedback for the TB of each stream to its serving gNB.

At the gNB side, a HARQ feedback (i.e., {DlHarqInfo} struct) from a UE is forwarded to the {NrMacSchedulerNs3} class for processing. In particular, we extended the {ProcessHarqFeedbacks} function of this class to read the HARQ feedback of each stream. Upon processing a HARQ feedback, if a UE has reported ACK for both streams' TB, the HARQ process id is released, i.e., it can be reassigned to a new DCI. However, it may happen that a UE is able to decode the TB of only one stream, i.e., it reports ACK for one stream and NACK for the other. In this case, while processing the feedback, the scheduler will not release the HARQ process id and reschedules the TB of the failed stream using the same HARQ process id. Moreover, the scheduler will prioritize retransmission of the TB of this single stream over scheduling new data during such retransmission. It will keep scheduling only one stream (for which it received NACK) until the TB is successfully decoded or the maximum number of retransmissions is reached.


\section{Example}
\label{sec:examples}
We created the {cttc-nr-mimo-demo.cc} example in the {nr} module to demonstrate the usage of the proposed MIMO framework. The topology consists of a single gNB and single UE. Simulation allows to configure various parameters, such as: the distance, the type of the RI value computation used (fixed or adaptive), the value of RI if fixed RI is used (1 or 2), the SINR thresholds if adaptive RI algorithm is used (see description in Sec.~\ref{sec:phy}), the random run number (to run multiple simulations and average the results), the 3GPP scenario, and the MCS table. The traffic used in the example is the downlink UDP constant bit rate (CBR). The output of the example, such as TX/RX bytes, throughput, mean jitter, and mean delay, is shown on the terminal and is written to a text file. 

\section{Simulation results}
\label{sec:results}
For the evaluation, we use ns-3 5G-LENA~\cite{PATRICIELLO2019101933} with the 3GPP spatial channel model developed in~\cite{tommaso:20}, compliant with TR 38.901, but extended to account for dual-polarized antennas, and the NR-based PHY abstraction model described in~\cite{lagen20}. 
The script used for evaluation is {cttc-nr-mimo-demo.cc}, explained in Sec.~\ref{sec:examples}. 

The scenario consists of 1 gNB and 1 UE, placed at a certain distance. The propagation condition follows the Urban Micro (UMi) scenario, 
as per TR 38.901~\cite{TR38901}. 
UMi is characterized by a gNB antenna height of 10~m, with a gNB transmit power of 30 dBm, and the UE has an antenna height of 1.5~m. 
The transmission is performed in the 3.5~GHz band region, using numerology 0 (i.e., 15~kHz subcarrier spacing), with 20~MHz channel bandwidth and a PRB overhead of 0.04 (typical of NR). 
The antenna array configuration consists of $2\times2$ dual-polarized directional antenna array at the gNB (i.e., the gNB has a total of 8 antenna elements) and $1\times1$ dual-polarized isotropic antenna array at the UE (i.e., the UE has a total of 2 antenna elements). We use MCS Table2 of NR that includes up to 256QAM.  With MCS Table2, 20~MHz channel bandwidth and numerology 0, the maximum achievable throughput results to be around 100 Mbps for 1 stream, and doubled ($\sim$200Mbps) for 2-streams' transmissions. Link adaptation is based on the Error model~\cite{lagen20}. For HARQ, we use Incremental Redundancy and up to 20 HARQ processes per UE.

The simulations are performed by varying the gNB-UE distance. We compare fixed RI scheme (RI equal to 1 or 2) and adaptive RI using {RiSinrThreshold1}= 7~dB and {RiSinrThreshold2}= 12~dB. Full buffer traffic with CBR and an UDP packet size of 1000~B is simulated. We consider UDP over RLC Unacknowledged Mode (UM). For each gNB-UE distance, results are averaged over 20 random channel realizations to get statistic significance. The end-to-end performance metric is the user throughput, at the IP layer.

\begin{figure}
    \centering
    \includegraphics[width=0.45\textwidth]{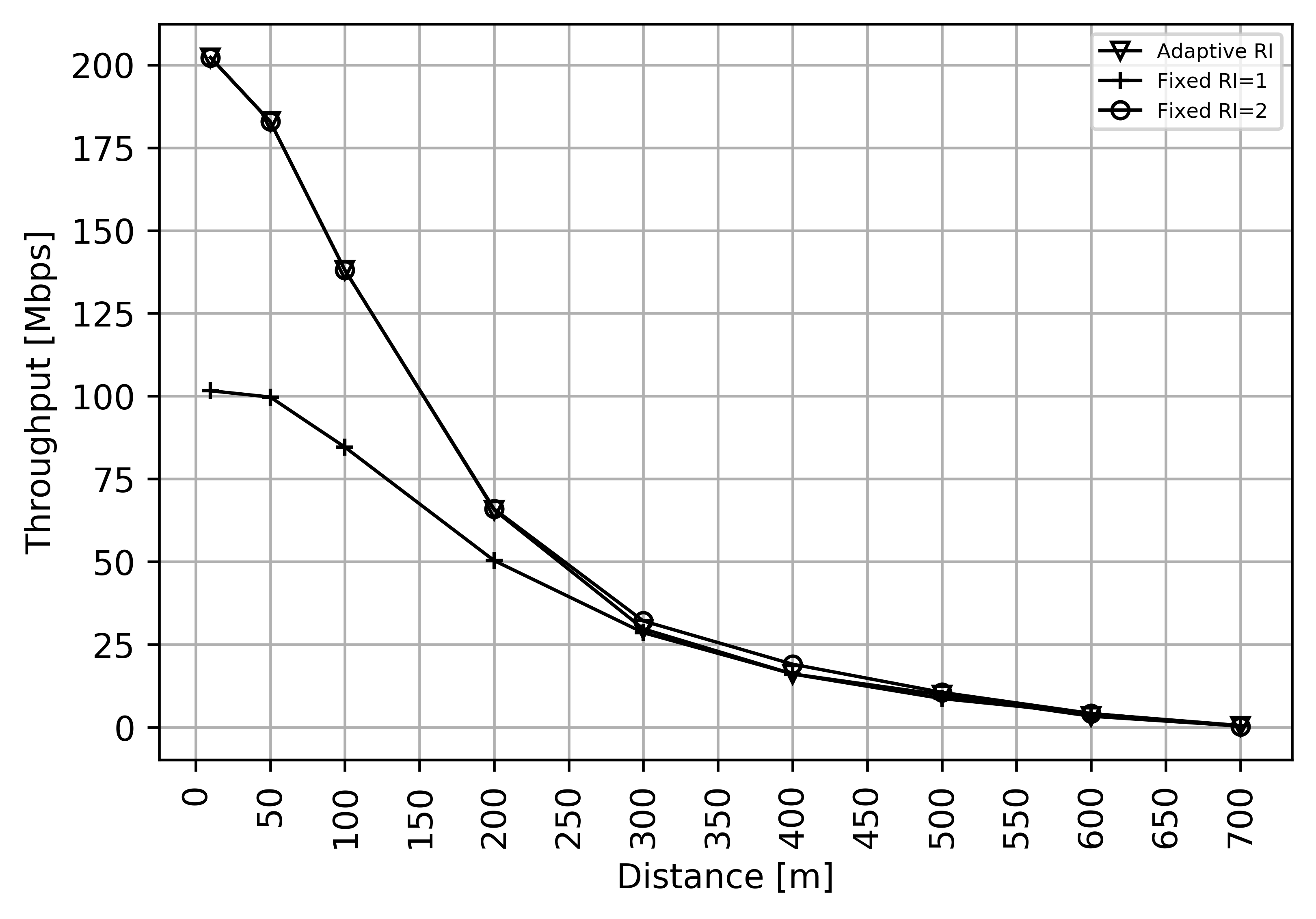}
    \caption{Throughput (Mbps) versus distance (m) for fixed RI (1 and 2) and adaptive RI algorithm}
    \label{fig:thput}
\end{figure}

Fig.~\ref{fig:thput} shows the average achieved throughput for different gNB-UE distances when using a fixed RI (i.e., fixed 1 or 2 streams) and the adaptive RI algorithm. As expected, for low distances, fixed RI=2 gets the maximum throughput of 200~Mbps and fixed RI=1 gets the half, i.e., 100~Mbps. The throughput starts to decrease when increasing the distance. So, the multiplexing gain is properly exhibited for good propagation conditions (i.e., low distances). As for the rank adaptation algorithm, we can see it also performs as expected: for low distances, RI=2 is chosen, because of the good propagation conditions on both the streams; while, as the distance increases, we can see that both RI=2 and RI=1 may be selected, depending on the specific simulation run, because the average throughput of the adaptive RI lies in between the fixed RI=2 and fixed RI=1 (see distance range from 300~m to 500~m in Fig.~\ref{fig:thput}). Finally, the throughput drops to 0 for very large distances in UMi scenarios, regardless the RI scheme.
These results validate the performance of fixed RI, as well as the proposed adaptive RI algorithm with two thresholds.


\section{Conclusions}
\label{sec:conclusions}
In this paper, we presented an extension of the ns-3 simulator and the 5G-LENA module to support DP-MIMO with spatial multiplexing of two streams. The developed MIMO model in the 5G-LENA exploits dual-polarized antennas to send two streams. 
The extension has implied major implementation changes in PHY and MAC layers of the {nr} module, as well as significant extensions in the ns-3 {spectrum} and {antenna} modules. We described the implementation changes and design choices in detail. Finally, we validated the developed DP-MIMO model in an Urban Micro scenario, for various gNB-UE distances, under a fixed rank (of 1 and 2) and the proposed rank adaptation algorithm.

\begin{acks}
This work was partially funded by Meta and Spanish MINECO grant TSI-063000-2021-56/TSI-063000-2021-57 (6G-BLUR). 
\end{acks}

\balance

\bibliographystyle{ACM-Reference-Format.bst}
\bibliography{wns3-2022.bib}


\newcommand{\SortNoop}[1]{}
\begin{thebibliography}{16}


\ifx \showCODEN    \undefined \def \showCODEN     #1{\unskip}     \fi
\ifx \showDOI      \undefined \def \showDOI       #1{#1}\fi
\ifx \showISBNx    \undefined \def \showISBNx     #1{\unskip}     \fi
\ifx \showISBNxiii \undefined \def \showISBNxiii  #1{\unskip}     \fi
\ifx \showISSN     \undefined \def \showISSN      #1{\unskip}     \fi
\ifx \showLCCN     \undefined \def \showLCCN      #1{\unskip}     \fi
\ifx \shownote     \undefined \def \shownote      #1{#1}          \fi
\ifx \showarticletitle \undefined \def \showarticletitle #1{#1}   \fi
\ifx \showURL      \undefined \def \showURL       {\relax}        \fi
\providecommand\bibfield[2]{#2}
\providecommand\bibinfo[2]{#2}
\providecommand\natexlab[1]{#1}
\providecommand\showeprint[2][]{arXiv:#2}

\bibitem[\protect\citeauthoryear{{3GPP}}{{3GPP}}{2019}]%
        {TR38901}
\bibfield{author}{\bibinfo{person}{{3GPP}}.} \bibinfo{year}{2019}\natexlab{}.
\newblock \bibinfo{title}{{Study on Channel Model for Frequencies from 0.5 to
  100 GHz}}.
\newblock \bibinfo{howpublished}{TR 38.901 (Rel. 15), V15.0.0}.
\newblock


\bibitem[\protect\citeauthoryear{{Bach Andersen}}{{Bach Andersen}}{2000}]%
        {842121}
\bibfield{author}{\bibinfo{person}{J. {Bach Andersen}}.}
  \bibinfo{year}{2000}\natexlab{}.
\newblock \showarticletitle{Antenna Arrays in Mobile Communications: Gain,
  Diversity, and Channel Capacity}.
\newblock \bibinfo{journal}{\emph{IEEE Antennas and Propagation Magazine}}
  \bibinfo{volume}{42}, \bibinfo{number}{2} (\bibinfo{date}{April}
  \bibinfo{year}{2000}), \bibinfo{pages}{12--16}.
\newblock


\bibitem[\protect\citeauthoryear{Baldo, Miozzo, Requena-Esteso, and
  Nin-Guerrero}{Baldo et~al\mbox{.}}{2011}]%
        {lena}
\bibfield{author}{\bibinfo{person}{N. Baldo}, \bibinfo{person}{M. Miozzo},
  \bibinfo{person}{M. Requena-Esteso}, {and} \bibinfo{person}{J.
  Nin-Guerrero}.} \bibinfo{year}{2011}\natexlab{}.
\newblock \showarticletitle{{An Open Source Product-oriented LTE Network
  Simulator Based on ns-3}}. In \bibinfo{booktitle}{\emph{Proceedings of the
  14th ACM International Conference on Modeling, Analysis and Simulation of
  Wireless and Mobile Systems}}. \bibinfo{address}{Miami, Florida, USA},
  \bibinfo{pages}{293--298}.
\newblock
\showISBNx{978-1-4503-0898-4}


\bibitem[\protect\citeauthoryear{Ben~Zid, Raoof, and Bouallegue}{Ben~Zid
  et~al\mbox{.}}{2012}]%
        {6206454}
\bibfield{author}{\bibinfo{person}{Maha Ben~Zid}, \bibinfo{person}{Kosai
  Raoof}, {and} \bibinfo{person}{Ammar Bouallegue}.}
  \bibinfo{year}{2012}\natexlab{}.
\newblock \showarticletitle{Dual polarized versus single polarized MIMO: A
  study over NLOS propagation with polarization discrimination and spatial
  correlation effects}. In \bibinfo{booktitle}{\emph{2012 6th European
  Conference on Antennas and Propagation (EUCAP)}}.
  \bibinfo{pages}{1979--1983}.
\newblock


\bibitem[\protect\citeauthoryear{Bojovic, Lagen, and Giupponi}{Bojovic
  et~al\mbox{.}}{2021}]%
        {bil21ns3}
\bibfield{author}{\bibinfo{person}{B. Bojovic}, \bibinfo{person}{S. Lagen},
  {and} \bibinfo{person}{L. Giupponi}.} \bibinfo{year}{2021}\natexlab{}.
\newblock \showarticletitle{{Realistic beamforming design using SRS-based
  channel estimate for ns-3 5G-LENA module}}. In
  \bibinfo{booktitle}{\emph{Proceedings of the Workshop on ns-3}}
  \emph{(\bibinfo{series}{WNS3 2021})}. \bibinfo{publisher}{Association for
  Computing Machinery}, \bibinfo{pages}{81--87}.
\newblock


\bibitem[\protect\citeauthoryear{Catreux, Greenstein, and Erceg}{Catreux
  et~al\mbox{.}}{2003}]%
        {1203169}
\bibfield{author}{\bibinfo{person}{S. Catreux}, \bibinfo{person}{L.J.
  Greenstein}, {and} \bibinfo{person}{V. Erceg}.}
  \bibinfo{year}{2003}\natexlab{}.
\newblock \showarticletitle{Some results and insights on the performance gains
  of MIMO systems}.
\newblock \bibinfo{journal}{\emph{IEEE Journal on Selected Areas in
  Communications}} \bibinfo{volume}{21}, \bibinfo{number}{5}
  (\bibinfo{year}{2003}), \bibinfo{pages}{839--847}.
\newblock


\bibitem[\protect\citeauthoryear{Dahlman, Parkvall, and Sköld}{Dahlman
  et~al\mbox{.}}{2018}]%
        {DAHLMAN20181}
\bibfield{author}{\bibinfo{person}{Erik Dahlman}, \bibinfo{person}{Stefan
  Parkvall}, {and} \bibinfo{person}{Johan Sköld}.}
  \bibinfo{year}{2018}\natexlab{}.
\newblock \bibinfo{booktitle}{\emph{5G NR: the Next Generation Wireless Access
  Technology}}.
\newblock \bibinfo{publisher}{Academic Press}.
\newblock
\showISBNx{978-0-12-814323-0}


\bibitem[\protect\citeauthoryear{{Femto Forum}}{{Femto Forum}}{[n.d.]}]%
        {ff_mac_sched}
\bibfield{author}{\bibinfo{person}{{Femto Forum}}.}
  \bibinfo{year}{[n.d.]}\natexlab{}.
\newblock \bibinfo{title}{LTE MAC Scheduler Interface Specification v1.11}.
\newblock \bibinfo{howpublished}{www.femtoforum.org}.
\newblock


\bibitem[\protect\citeauthoryear{Ikuno, Pendl, Šimko, and Rupp}{Ikuno
  et~al\mbox{.}}{2012}]%
        {6364098}
\bibfield{author}{\bibinfo{person}{Josep~Colom Ikuno}, \bibinfo{person}{Stefan
  Pendl}, \bibinfo{person}{Michal Šimko}, {and} \bibinfo{person}{Markus
  Rupp}.} \bibinfo{year}{2012}\natexlab{}.
\newblock \showarticletitle{{Accurate SINR estimation model for system level
  simulation of LTE networks}}. In \bibinfo{booktitle}{\emph{2012 IEEE
  International Conference on Communications (ICC)}}.
  \bibinfo{pages}{1471--1475}.
\newblock


\bibitem[\protect\citeauthoryear{Lag\'{e}n, Wanuga, Elkotby, Goyal,
  Patriciello, and Giupponi}{Lag\'{e}n et~al\mbox{.}}{2020}]%
        {lagen20}
\bibfield{author}{\bibinfo{person}{Sandra Lag\'{e}n}, \bibinfo{person}{Kevin
  Wanuga}, \bibinfo{person}{Hussain Elkotby}, \bibinfo{person}{Sanjay Goyal},
  \bibinfo{person}{Natale Patriciello}, {and} \bibinfo{person}{Lorenza
  Giupponi}.} \bibinfo{year}{2020}\natexlab{}.
\newblock \showarticletitle{{New Radio Physical Layer Abstraction for
  System-Level Simulations of 5G Networks}}. In
  \bibinfo{booktitle}{\emph{Proceedings of IEEE International Conference on
  Communications}} \emph{(\bibinfo{series}{IEEE ICC})}.
  \bibinfo{address}{Virtual Conference}.
\newblock


\bibitem[\protect\citeauthoryear{Mezzavilla, Zhang, Polese, Ford, Dutta,
  Rangan, and Zorzi}{Mezzavilla et~al\mbox{.}}{2018}]%
        {mezzavilla2017end}
\bibfield{author}{\bibinfo{person}{Marco Mezzavilla}, \bibinfo{person}{Menglei
  Zhang}, \bibinfo{person}{Michele Polese}, \bibinfo{person}{Russell Ford},
  \bibinfo{person}{Sourjya Dutta}, \bibinfo{person}{Sundeep Rangan}, {and}
  \bibinfo{person}{Michele Zorzi}.} \bibinfo{year}{2018}\natexlab{}.
\newblock \showarticletitle{{End-to-End Simulation of 5G mmWave Networks}}.
\newblock \bibinfo{journal}{\emph{IEEE Communications Surveys \& Tutorials}}
  \bibinfo{volume}{20}, \bibinfo{number}{3} (\bibinfo{date}{April}
  \bibinfo{year}{2018}), \bibinfo{pages}{2237--2263}.
\newblock


\bibitem[\protect\citeauthoryear{Patriciello, Lag\'{e}n, Bojovi\'{c}, and
  Giupponi}{Patriciello et~al\mbox{.}}{2019a}]%
        {PATRICIELLO2019101933}
\bibfield{author}{\bibinfo{person}{Natale Patriciello}, \bibinfo{person}{Sandra
  Lag\'{e}n}, \bibinfo{person}{Biljana Bojovi\'{c}}, {and}
  \bibinfo{person}{Lorenza Giupponi}.} \bibinfo{year}{2019}\natexlab{a}.
\newblock \showarticletitle{{An E2E Simulator for 5G NR Networks}}.
\newblock \bibinfo{journal}{\emph{Simulation Modelling Practice and Theory}}
  \bibinfo{volume}{96} (\bibinfo{date}{Nov.} \bibinfo{year}{2019}),
  \bibinfo{pages}{101933}.
\newblock
\showISSN{1569-190X}


\bibitem[\protect\citeauthoryear{Patriciello, Lagen, Giupponi, and
  Bojovic}{Patriciello et~al\mbox{.}}{2019b}]%
        {NatlieMac}
\bibfield{author}{\bibinfo{person}{N. Patriciello}, \bibinfo{person}{S. Lagen},
  \bibinfo{person}{L. Giupponi}, {and} \bibinfo{person}{B. Bojovic}.}
  \bibinfo{year}{2019}\natexlab{b}.
\newblock \showarticletitle{{An Improved MAC Layer for the 5G NR ns-3 module}}.
  In \bibinfo{booktitle}{\emph{WNS3 2019, June 2019, Firenze, Italy}}.
\newblock


\bibitem[\protect\citeauthoryear{Pi and Khan}{Pi and Khan}{2011}]%
        {pi:11}
\bibfield{author}{\bibinfo{person}{Zhouyue Pi} {and} \bibinfo{person}{Farooq
  Khan}.} \bibinfo{year}{2011}\natexlab{}.
\newblock \showarticletitle{An Introduction to Millimeter-Wave Mobile Broadband
  Systems}.
\newblock \bibinfo{journal}{\emph{IEEE Communications Magazine}}
  \bibinfo{volume}{49}, \bibinfo{number}{6} (\bibinfo{date}{June}
  \bibinfo{year}{2011}), \bibinfo{pages}{101--107}.
\newblock


\bibitem[\protect\citeauthoryear{Qin, Guo, and Liang}{Qin
  et~al\mbox{.}}{2010}]%
        {5638598}
\bibfield{author}{\bibinfo{person}{Pei-Yuan Qin}, \bibinfo{person}{Y.~Jay Guo},
  {and} \bibinfo{person}{Chang-Hong Liang}.} \bibinfo{year}{2010}\natexlab{}.
\newblock \showarticletitle{Effect of Antenna Polarization Diversity on MIMO
  System Capacity}.
\newblock \bibinfo{journal}{\emph{IEEE Antennas and Wireless Propagation
  Letters}}  \bibinfo{volume}{9} (\bibinfo{year}{2010}),
  \bibinfo{pages}{1092--1095}.
\newblock


\bibitem[\protect\citeauthoryear{Zugno, Polese, Patriciello, Bojovi\'{c},
  Lag\'{e}n, and Zorzi}{Zugno et~al\mbox{.}}{2020}]%
        {tommaso:20}
\bibfield{author}{\bibinfo{person}{Tommaso Zugno}, \bibinfo{person}{Michele
  Polese}, \bibinfo{person}{Natale Patriciello}, \bibinfo{person}{Biljana
  Bojovi\'{c}}, \bibinfo{person}{Sandra Lag\'{e}n}, {and}
  \bibinfo{person}{Michele Zorzi}.} \bibinfo{year}{2020}\natexlab{}.
\newblock \showarticletitle{{Implementation of a Spatial Channel Model for
  ns-3}}. In \bibinfo{booktitle}{\emph{Proceedings of the 2020 Workshop on
  ns-3}}. \bibinfo{pages}{49--56}.
\newblock


\end{thebibliography}
\end{document}